\documentstyle[11pt]{article}
\begin{document}

\title{Possible tests of curvature effects in weak gravitational fields}

\author{Bruno Preziosi and Guglielmo M. Tino\and
Dipartimento di Scienze Fisiche, Universit\`a di Napoli "Federico II"\and
Napoli, Italy}

\maketitle

\begin{abstract}

We analyze the possibility of detecting, with optical methods,
particle and photon trajectories predicted by general relativity 
for a weak spherically-symmetric 
gravitational field.
We discuss the required sensitivities and the possibility of
performing specific experiments on the Earth. 
\end{abstract}


of gravitational radiation.

06.30 - Measurement of basic quantities.\\

\vskip 1cm

mailing address:\\
Prof. Bruno Preziosi\\
Dipartimento di Scienze Fisiche, Universit\'a di Napoli "Federico II"\\
Mostra d'Oltremare, Pad.20\\
I-80125 Napoli (Italy)\\
tel.n.: (+39) 81-7253416\\
e-mail: preziosi@na.infn.it\\
fax n.: (+39) 81-2394508\\

\newpage

\section{Introduction}
\label{s1}

Since the formulation of general relativity (GR),
several experiments have been proposed and/or performed
in order to test the predictions of this theory \cite{review}.
The most famous GR tests are concerned with the deflection of
light by the Sun's gravitational field, with the
perihelion shift of Mercury and, more recently,  with the radar echo delays
\cite{Shap71,Reas79}.

On-Earth experimental tests have also provided a richness of information,
although they are
extremely difficult because of the small size of the effects to
be observed.
The precision achievable in different experiments
has been largely increased by the use of maser and laser sources.
This accuracy was high enough to perform, for example, stringent
tests of the isotropy of the light velocity \cite{Bril79}.
Similar techniques have been used to improve the
gravitational red shift experiment, first performed 
by Pound and Rebka \cite{Poun60},
by comparing the frequency of a hydrogen maser in a ground laboratory 
with that of an identical maser on board of a rocket \cite{Ves80}.
Moreover, several experiments have been recently performed to search
for a "fifth-force" \cite{Nie87}. Such experiments provide an upper
limit to a violation of the Einstein equivalence principle.
Matter wave interferometers have also been used to study the influence
of the homogeneous Earth gravitational field on quantum systems and
the measurability of the influence of the space-time curvature
was recently discussed \cite{Aud94}.

In this paper we analyze the possibility of performing, on or near the 
Earth, experiments of the
type based on the space curvature induced by the Sun.
We discuss in particular three possible "local" experiments based on optical
techniques to search for space-time curvature effects 
on the Earth or in its proximity, taking advantage 
of the extremely high sensitivity achievable with 
available optical techniques. On-Earth experiments offer
the advantage of a better control of the experimental parameters. 
 The first experiment we discuss is aimed to the detection
of tidal acceleration using a freely-falling Michelson interferometer. 
The basic idea of the second experiment is instead to put into 
evidence, using laser ranging techniques,
the relativistic correction, at the second order in the Schwarzschild
radius, to the classical gravity acceleration.
 We investigate finally the possibility of detecting light deflection 
induced by space curvature by following its propagation in an optical 
cavity made of two high-reflectivity mirrors.

Although the purpose of this paper is not
to design real experiments, we discuss the possibility
of performing them and the sensitivities
required in order to detect the relevant effects.

The paper is organized as follows: in Sect. 2, we 
give, with a relatively simple procedure, the 
expression for the trajectories of photons and of massive
bodies in a weak Schwarzschild field \cite{Schwarz}. 
This allows us to calculate the size of the relevant effects 
in the following discussion of possible experimental tests.
In Sect. 3, we present the idea of using a falling
Michelson interferometer to measure the tidal acceleration.
In Sect. 4, we consider a possible scheme
to measure the correction to $g$ due to the space curvature.
In Sect. 5, we discuss the possibility of a local detection of the 
light deflection due to the Earth gravitational field.
In Sect. 6 we draw final conclusions about the 
feasibility or not of each of the experiments.

\section {Light and particle motion in a Schwarzschild field}
In this section we give the main results of a calculation 
of the trajectory of light and of massive particles in
a weak spherically-symmetric gravitational field.
Only the results which are important for the experiments 
discussed in the following are reported. 
A more detailed account of this calculation will be given elsewhere \cite{Pre}.

First, we consider photons.
The equations for the null geodesics in a Schwarzschild field, described by
the metric
 \begin{equation}
 ds^{2}=-\left(1-\frac{2K}{r}\right)c^{2}dt^{2}~+~\frac{1}{1-\frac{2K}{r}}dr^{2}
 +r^{2}(\sin^{2}\theta d\phi^{2}+d \theta^{2})
 \end{equation}
 where $K=GM/c^{2}$ is the Schwarzschild radius, when expressed in terms
of a suitable parameter $\tau$,
are the following: \cite{Chan83,Papa74}
\begin{equation}
\left(\frac{dr}{d\tau}\right)^{2}+\frac{L^{2}}{r^{2}}\left(1-\frac{2K}{r}
\right)=E^{2},~~~~~c\frac{dt}{d\tau}=\frac{E}{1-2K/r},~~~~~
\frac{d\theta}{d\tau}=\frac{L}{r^{2}},
\end{equation}
where $r$ and $t$ are the Schwarzschild variables and $L$ is $c$ times 
the angular momentum. 
The trajectories seen by a stationary observer in the planet
reference frame are expressed in terms 
of the local variables 
\cite{Misn73,Light75,Ken90}  
$\hat{r}, \hat{t}$ and $\hat{\theta}$, which 
are obtained from the espressions of
$r, \theta$ and $t$ along the trajectories from:
\begin{equation}
d\hat{r}\breve{=}\frac{dr}{\sqrt{1-2K/r}},~~~~~
\hat{r}d\hat{\theta}\breve{=}rd\theta,~~~~~d\hat{t}\breve{=}dt\sqrt{1-2K/r}.
\end{equation}
The symbol $\breve{=}$ is used to remark that its right hand 
side must be expressed
in terms of the Schwarzschild solution.
Our initial conditions are such that, at $\tau=t=\theta=\hat{t}=\hat{\theta}
=0$, the photon moves
through a point at a distance
$\hat{r}=R$ from the center of the planet  along a direction which makes
an angle $\alpha$ with the radial one; the local initial motion is then:
\begin{equation}  
\hat{r}=R\frac{\sin\alpha}{\sin(\alpha-\hat{\theta})};
~~~~L=ER\sin\alpha
\end{equation}
The procedure to derive the local motion is the following:
first of all, we derive, from eqs. (2), the expressions for $r,\theta,t$
at the fourth order in $\tau$ and at the second order in $K$; then, from
eqs. (3), we derive the analogous polinomials for $\hat{r},\hat{\theta}$
and $\hat{t}$, which satisfy the given initial conditions. These solutions,
when expressed 
in terms of local Cartesian coordinates $\hat{z}\equiv\hat{r}\cos\hat{\theta}
-R$ and $\hat{x}\equiv\hat{r}\sin\hat{\theta}$ assume the following form:
\begin{eqnarray}
& &\hat{z}=c\hat{t}\cos\alpha-\frac{1}{2}\frac{K}{R}\sin^{2}\alpha
\left[\left(1+3\frac{K}{R}\right)\frac{c^{2}\hat{t}^{2}}{R}-
\left(\frac{4}{3}+\frac{11K}{3R}\right)\cos\alpha
\frac{c^{3}\hat{t}^{3}}{R^{2}}\right]
+\nonumber\\
& &+\frac{1}{3}\frac{K}{R}\sin^{2}\alpha
\left[\frac{1}{2}-\frac{23}{8}\cos^{2}\alpha
+\left(1
-6\cos^{2}\alpha
\right)\frac{K}{R}\right]\frac{c^{4}\hat{t}^{4}}{R^{3}}
\end{eqnarray}
\begin{eqnarray}
& &\hat{x}=c\hat{t}\sin\alpha+\frac{1}{2}\frac{K}{R}\sin\alpha
\cos\alpha\left(1+3\frac{K}{R}\right)\frac{c^{2}\hat{t}^{2}}{R}+\nonumber\\
& &-\frac{1}{3}\frac{K}{R}\sin\alpha
\left[2\cos^{2}\alpha
+\left(\frac{1}{2}+\frac{11}{2}\cos^{2}\alpha\right)\frac{K}{R}
\right]\frac{c^{3}\hat{t}^{3}}{R^{2}}+\\
& &+\frac{1}{6}\frac{K}{R}\sin^{2}\alpha\cos\alpha
\left[-1+\frac{23}{4}\cos^{2}\alpha
+\left(1+12\cos^{2}\alpha\right)\frac{K}{R}\right]
\frac{c^{4}\hat{t}^{4}}{R^{3}}\nonumber
\end{eqnarray}
It is interesting to notice that, for $\alpha=\pi/2$, the trajectory in 
the Schwarzschild variables,
which corresponds to the previous solution and solves eqs. (2), is such that:
\begin{equation}
r^{-1}\simeq\frac{\cos(\theta)}{R}+\frac{K}{R^{2}}(2-\cos^{2}\theta)
\end{equation}
As a consequence,  $r^{-1}\rightarrow 0$
when $\theta \rightarrow \pm\frac{\pi}{2}\pm\frac{2K}{R}$. The well-known
expression $4K/R$ for light deviation follows.

For the following discussion, it is also important to calculate the motion 
of massive particles.
The equations governing the radial geodesics, in units $c=1$,
are \cite{Chan83}:
\begin{equation}
\left(\frac{dr}{d\tau}\right)^{2}+(1-\frac{2K}{r})=E^{2},~~~~~~~
\frac{dt}{d\tau}=\frac{E}{1-2K/r}.
\end{equation}
where $r$ and $t$ are the Schwarzschild variables.
The same procedure used for the photon case can be followed.
First of all, one looks for the generic polynomials at the fourth order in 
$\tau$ and at the second order in $K$ which solve equations (8).
Then one finds the similar polynomials in $\tau$
which give $\hat{r}$ and $\hat{t}$.
Finally one fixes the arbitrary parameters in such a way that 
the initial conditions ($\hat{r})_{\hat{t}=0}=R, (d\hat{r}/d\hat{t})_{\hat{t}=0}
=v_{0}$ are satisfied.
Re-introducing $c$, the final result is the following:

\begin{eqnarray}
& &\hat{r}=R+v_{0}\hat{t}-\frac{1}{2}\frac{K}{R}
\left(1-\frac{v_{0}^{2}}{c^{2}}\right)
\left[\left(1+3\frac{K}{R}\right)
\frac{c^{2}\hat{t}^{2}}{R}-\frac{v_{0}}{c}
\left(\frac{2}{3}+\frac{5K}{3R}\right)
\frac{c^{3}\hat{t}^{3}}{R^{2}}\right]+\nonumber\\
& &-\frac{1}{4}\frac{K}{R}\left[1-\frac{v_{0}^{2}}{c^{2}}\right)^{2}
\left(\frac{v_{0}^{2}}{c^{2}-v_{0}^{2}}
+\frac{1}{3}\left(1+5\frac{v_{0}^{2}}{c^{2}-v_{0}^{2}}\right)
\frac{K}{R}\right]
\frac{c^{4}\hat{t}^{4}}{R^{3}}.
\end{eqnarray}

Notice that, in the limit $v_{0}\rightarrow c$, $\hat{r}\rightarrow
R+ct$. If the particle is initially at rest, the previous expression
reduces to
\begin{equation}
\hat{r} \simeq R-\frac{1}{2}g\hat{t}^{2}-\frac{1}{12}\frac{g^{2}}{R}\hat{t}^{4},
\end{equation}
where 
\begin{equation}
g \equiv \frac{GM}{R^{2}}\left(1+3\frac{GM}{Rc^{2}}\right).
\end{equation}
We verified that expression (11) for the gravity acceleration
remains unchanged at the same level of approximation also if the Kerr metric
is used to take into account possible effects induced by the rotation 
of the gravitational field generator.

The trajectory of a photon,
initially moving through $O$ perpendicularly to the radius ($\alpha=\pi/2$),
is given by (see eqs (5) and (6)):
\begin{eqnarray}
\hat{x}\simeq c\hat{t}-\frac{1}{6}\frac{g^{2}}{c}\hat{t}^{3},\\
\hat{z}\simeq -\frac{1}{2}g\hat{t}^{2}+\frac{1}{6}g\hat{t}^{2}
\frac{c^{2}\hat{t}^{2}}{R^{2}}+\frac{1}{3}\frac{g^{2}}{R}\hat{t}^{4}.
\end{eqnarray}

Notice that the light velocity along its trajectory is constant.
It is obvious that the above expansion is valid only if
 the distance travelled by the light, which differs from the
Pythagoric distance only at the second order in $g$, 
is much shorter than the distance from the planet center. 
The last term in eq. (12), which coincides with the classical correction
due to the non uniformity of the gravitational field, and the last term
in eq. (13) will be neglected in the next as they are too small. The second
term at the r. h. s. of eq. (13) is apparently due to a repulsive force; 
the reason for this is that, when the light beam moves in the horizontal
direction, it goes towards regions with decreasing gravity.
In the following we will omit the sign $\hat{}$ so that $x,z$ and
$t$ will indicate the local variables.

\vspace{1cm}

\section {Measurement of the tidal effect}

In this section we analyze the possibility of detecting tidal acceleration 
effects, due to the non-uniformity of the gravitational field.

We consider a freely-falling Michelson interferometer characterized by
 the following scheme (see Fig. 1):
inside an {\it elevator} a beam splitter $O$ and two
mirrors $A$ and $C$ are placed in the horizontal plane; $A$ is freely falling
while $C$, $O$, the laser source $S$ and a photodetector $B$  are rigidly
connected and freely falling.
We suppose that the distances $OC$ and $OA$ are initially equal.
In the horizontal plane $(x,y)$  the initial positions of
the componentss are the following: $O\equiv(0,0),A\equiv(D,0), 
C\equiv(0,D), S\equiv (-d,0), B\equiv (0,-d)$; ($d \ll D$).
Neglecting, for the moment, higher order terms
in the  frame
in which $O$ and
the light beam move like in the uniform case, the mirror $A$
moves along the geodesics:
\begin{equation}
x=L-\frac{1}{2}gt^{2}\sin\theta;~~~~~~~z=-\frac{1}{2}gt^{2}\cos\theta,
\end{equation}
with $\sin\theta=D/R$.
From the point of view of the frame in which
$O$ is stationary,  neglecting terms of the
order $D^{2}/R^{2}$, $A$ is moving toward $O$
according to the law:
\begin{equation} 
x'\simeq D-\frac{1}{2}gt'^{2}\sin\theta\simeq D-\frac{1}{2}g\frac{D}{R}
t'^{2};~~~~~~z'=0;
\end{equation}
if the light beam to $A$ leaves $O$ at the time $\tilde{t}'$, 
it will return in $O$
at the time $\tilde{t}'+2\frac{D}{c}- \frac{D}{R}\frac{g\tilde{t}'^{2}}{c}$.
Neglecting the frequency variation due to the mirror motion, the phase
difference between the beam from $A$ and the one from $C$ is given by
$2 \pi \nu'\frac{D}{cR}g\tilde{t}'^{2}$. The phase variation rate
is then given by $2 \pi\frac{D}{R}\frac{g\tilde{t}'}{\lambda'}$;
in these expressions
$\nu' (\lambda')$ is the laser source frequency (wavelength), 
which is seen by $\Sigma^{'}$ to be 
constant. 
The phase change that would be observed for an interferometer falling
in proximity of the Earth can be evaluated considering visible light
$(\lambda'= 0.6 \mu m)$, $D$ = 1 m, $R = 6.4$x$10^{6}$ m. 
This gives a phase difference numerically given by $15 t'^{2}$,
where $t'$ is expressed in seconds. A large effect 
would then be observed in experiments with 
a free-fall time of only a few seconds.

Obviously, in a real experiment several experimental details should 
be accurately considered. First of all the experiment must be performed in
a region in which gravity acceleration is not affected by local
non-uniform mass distributions.
A critical point is the possibility of releasing the apparatus
while keeping the interferometer well aligned. The optical components
can be kept in place by means of small electro-magnets. The whole apparatus
is then released and the magnets are switched off. The available time should
be long enough to 
allow a slow release of the components in order to avoid spurious 
inductive effects. In fact, similar problems have been
already considered in the development of gravitometers based on 
an interferometer with
a falling mirror \cite{Nie87}. The use of retroreflectors (corner cubes),
for example, reduces the sensitivity to small misalignments.
A  vacuum chamber is required in order to
reduce the effect of air resistance to a negligible level.\\
As in the modern version of the Michelson-Morley experiment \cite{Bril79}, 
a large improvement in the sensitivity could be obtained
by using two orthogonal 
optical cavities instead of
a Michelson interferometer.

\vspace{1cm}

\section {Corrections to g}

This section is devoted to analyze the possibility of detecting 
GR effects in the vicinity of the Earth,
which is supposed, for the moment, to be spherical (the eventual
corrections will be discussed at the end of this section).

Let us suppose that  very accurate measurements of the gravity acceleration
in several points on the same radial direction at distances $R$ and $R+z_{i}$ 
from the center of the planet give the values $g$ and $g_{i}$ respectively. 
It is then possible, using eq. (11),  to obtain values for
$GM$ and for the reference distance $R$ with the same accuracy and to
 test the correctness of the motion
laws (10) and eventually (9); in particular one might control
whether or not the classical relation between the gravity 
acceleration and the gravitational constant is affected by the small correction 
$3\frac{GM}{Rc^{2}}$ (of the order of $2\cdot10^{-9}$). 
Notice that this correction is of the second order in $K/R$, while
the classical GR experiments are of the first order.
In principle, one might obtain gravity measurements using the
gravitometer as the one realized by Niebauer et al. \cite{Nie87}  which has the
required accuracy, but its use is forbidden by the requirement of
performing these measurements at very large distances.
For this reason, in the following, $g_{i}$ will be deduced from
laser ranging measurements of the distances between the falling body and 
an observation
point. This observation point may be placed on the Earth, in which case,
although one is interested in distances between positions which are
outside the atmosphere, the refraction and turbolence effects need to be
considered. In alternative, the observation points will be supposed 
on the Moon or on two satellites. We discuss in the next the latter case.

Let C be a reflecting  body, which is freely falling from an height of
$\sim 8000 Km$ from the center of the Earth in the plane $(x,z)$.
For the moment, we suppose that
the orbital angular momentum of $C$ is absolutely negligible and neglect
the Moon attraction.
 Let $A$ and $B$ be two satellites  which continuously
control their relative separation. A convenient (of the order of 100)
number of $\sim 1~ ps$ light 
pulses are sent  each $\sim 10^{-6} s$ from $A$ 
 towards $C$ and
times are taken when the reflected light reaches both 
$A$ and $B$; in this way
the relatives distances are known with a precision better than 
$10^{-9}$, if the relative distances are of the order of $1,500 Km$.
If ${t_{1}}$ is the mean value  of the times in which these flashes 
reached $C$, its positions could be approximated by
eq. (9) with the substitutions $\hat{t}\rightarrow t-{t_{1}}, 
R\rightarrow {R_{1}},
v_{0}\rightarrow {v_{1}}, K/R\rightarrow k_{1}',
 K^{2}/R^{2}\rightarrow k_{1}''$; during this  data acquisition time, 
we suppose that the polar coordinates
of $A$ and $B$, if not yet known,  change according to the laws
$(r_{A}+\dot{r}_{A}(t-t_{1}),\theta_{A}+\dot{\theta}_{A}(t-t_{1})_,\varphi_{A}+
\dot{\varphi}_{A}(t-t_{1})),
(r_{B}+\dot{r}_{B}(t-t_{1}),\theta_{B}+
\dot{\theta}_{B}(t-t_{1}),\varphi_{B}+\dot{\varphi}_{B}(t-t_{1})$,
which must be used also for the time data relativistic corrections.
A best fitting of the
 measured and the calculated distances will give the values of these 
parameters at $t=t_{1}$.

This procedure is repeated several times with a suitable rate.
If not only the products $k_{m}'R_{m}$ are equal to a constant $K$,
but also the products $k_{m}''R_{m}^{2}$  are constant and equal 
to $K^{2}$, then we may conclude that
effectively the relation between $g$ and $G$ has the form (11).

The non-sphericity of the Earth implies that the expansion of $g$ 
contains also latitude-dependent powers of $R^{-2}$ \cite{Heis58}, 
which can be taken into account by generalizing eq. (11). Moreover account
must be taken for the Moon presence which, near the Earth, gives a constant 
contribution to the acceleration.

For these reasons, it might be considered the possibility of using for $A,B$
and $C$ three objects which are orbiting around the Sun, as in this case we
do not need to consider corrections of the type described above.
This offers the advantage of operating in
 regions in which $K/R>2\cdot 10^{-9}$ (for example, 
at the level of our distance from the Sun, $K_{Sun}/R\approx 10^{-8}$).

The generalization of the method to the case in which $C$ has an
initial orbital angular momentum is obtained 
by multiplying the second term at the l.h.s. of the eq. (8)
by $(1+L^{2}/r^{2})$ and by considering also the last of eqs. (2); then one
finds the polynomials which solve these equations
at the fourth order in $t$ and at the second order in $K/R$. Following the
procedure described in section 2 one finds the final, rather complicated,
expressions which
give $\hat{r}$ and $\hat{\theta}$ in terms of $\hat{t}$; these expressions
contain also the GR corrections at the first order in $K/R$, analogous
to the one which enter in the light deflection from a Schwarzschild field
(see eqs (7), (6) and (5)).
For a possible experiment a computational solution will suffice.

\vspace{1cm}

\section{ Light deflection in a Schwarzschild field}

In analogy with the GR test on the light deflection by the Sun,
we  discuss in this section the possibility of locally detecting the light
deflection due to the Earth gravitational field.
The deflection is in this case extremely small; 
for a path of 5000 Km it is only  $\approx 2\cdot 10^{-9} rad$.

For this purpose we consider the light propagation in 
an optical cavity made of two high-reflectivity mirrors.
Optical cavities with a length of the order of a meter and
with finesse in the range $10^{5}-10^{6}$ can already be realized 
with present technology \cite{Hall94,DeR95}.
We consider here cavities with a finesse of the order of 
the ones presently achievable, the distance between
the mirrors being of the order of 100 m. 

We first  consider a cavity made of two plane mirrors;
such a configuration is not appropriate in a real experiment,
because of the sensitivity to misalignment and because of the unavoidable 
divergence of the laser beams which would mask any effect due to gravity.
However, its simplicity allows
to get a first insight into the relevant effects.  
We then discuss the more realistic case of stable 
cavities including non-planar mirrors.

Let us follow the motion of a photon which
enters the cavity, at $t=0$, 
going in the horizontal direction.
For a mirror reflectivity $r_{e}$ such that
$1-r_{e}=10^{-5}$ and for a distance between the mirrors $2d\simeq 100~m$,
the cavity optical decay time 
is 30 $ms$. 
If we consider only the first term in eq.(13), that is if
$z=z_{o}-\frac{1}{2}gt^{2}$,
the corresponding vertical
displacement of the beam is $5~mm$.
If we include the second non-uniform term of eq.(13), that is, if 
$z=z_{o}-\frac{1}{2}gt^{2}\left(1-\frac{1}{3}
\frac{c^{2}t^{2}}{R^{2}}\right)$, 
the photon fall is slightly retarded. 
For time intervals of the order of $30~ ms$ 
this expansion is not appropriate and further terms should be included,
but it suffices for the transit time between the two mirrors; the photon
motion is then correctly described by eqs. (5) and (6) if, after any
reflection, the new initial conditions are introduced.
 
As already mentioned, the configuration to be 
considered for a real experiment would be a cavity with 
 mirrors of spherical type.
This allows, with a proper choice of 
the curvature of the mirrors and of the distance between them,
to control the light beam divergence.
In order to analyze if, in a real experiment, it is possible to
see this beam displacement, one may calculate, using eqs. (5) and (6),
 the trajectory followed 
by the photons while travelling between two 
mirrors  of appropriate form. For brevity, we do not give here details on these
calculations.

 The result indicates that, if the two mirrors have spherical form 
any gravity
 effect is cancelled. 
This result, which is in contrast with what one might have expected 
considering the simple flat-mirrors model discussed above, 
excludes the possibility of detecting
gravity effects in typical very-high-Q optical cavities.
The qualitative explanation
 for this is that any photon which tends to go downwards because
of the beam divergence (or the gravity deflection) is reflected by the
cavity mirrors in the opposite direction.
Incidentally, this conclusion allows to exclude the effect of light deflection
in those experiments, such as gravitational wave detection, in which
very-high-Q optical cavities are used.

 A possibility of recovering the effect suggested by the plane case is to
take mirrors with a more complex shape.
We have analyzed the behaviour of a light beam, with
a Gaussian distribution characterized by a spread in $\varphi$ 
$(\simeq 10^{-4}-10^{-5})$
much smaller than the spread in $\zeta$, in
the case the two mirrors are essentially
made of a central cylindrical  part (of the order of $1~ cm$) with 
spherical confocal concavities at the upper and lower borders.
The conclusion of this analysis is
 that the beam motion appears to be partially chaotic 
and only a qualitative evidence of the light deflection might be reached
when the flat parts of the mirrors were parallel.  
The problem is that the experimental optimization of the cavity would be 
realized
not by this configuration, but when the mirrors
form an angle equal to the angular deviation of the photon
in going from the center of a mirror to the center of the other one;
this angle is equal to $2gd/c^{2}$.
As a consequence, the photon at the center of the beam moves from one
of the mirrors in the direction perpendicular to it and hits the 
other one perpendicularly; it remains then, after any reflection, on
the same geodesics \cite{Larocca}. 
If this is the case, the cavity is optimized and no
beam fall could be detected.

However, if the experiment were performed with a shorter and mechanically
rigid cavity, a rotation of the cavity around the longitudinal axis,
should show the resulting misalignment. 
Another possibility is to align the cavity and then let it
fall freely. In this case, the effect of the space curvature would
produce a deformation of the eigen-modes of the cavity. A quantitative
calculation of this effect would then require an analysis of the 
propagation of a light wave in a curved space-time,  but this is beyond
the scope of this work.

\vspace{1cm}

\section{Conclusions}

We described effects of a weak
spherically-symmetric gravitational field which might be
detected exploiting the high sensitivity achievable with laser sources
and modern optical methods.

Three possible experiments near the Earth were discussed. The result of our
 analysis seems to exclude the 
possibility of detecting the light deflection in an optical cavity. It
seems instead possible, using Michelson interferometers, to detect the
tidal acceleration. Finally, we proposed an experiment 
to measure GR corrections for $g$ on the Earth or in its
proximity; this correction can be put into evidence by
comparing the motion of a falling body 
with the Schwarzschild local solutions.

\vspace{.5cm}

{\bf Acknowledgments}

The authors thank M. Inguscio, G. Marmo and G. La Rocca for stimulating
discussions and useful suggestions.

\newpage

{\bf Figure captions}

\vspace{0.5cm}

{\bf Fig.~1.} Schematic diagram of the falling Michelson 
interferometer. The mirror A is falling freely. The source 
S, the beam splitter O, the mirror C and the detector B are
connected together and freely-falling.

\end{document}